\documentclass[reqno,11pt]{amsart}
\usepackage{amsfonts, amsthm}

{ \theoremstyle{definition}
 }

\usepackage{hyperref}

\begin{document}

\title[]{The degeneracy of Laplace invariants for hyperbolic systems possessing integrals}
\author[]{S. Ya. Startsev}

\urladdr{\href{http://www.researcherid.com/rid/D-1158-2009}{http://www.researcherid.com/rid/D-1158-2009}}

\address{Institute of Mathematics, Ufa Scientific Center, Russian Academy of Sciences}

\begin{abstract}
A direct generalization of Laplace invariants to the case of hyperbolic partial differential systems is considered. The proof of the following statement is given: the determinant of a Laplace invariant vanishes if the corresponding system admits an integral.   
\end{abstract}



\maketitle

The Laplace method of cascade integration is a classical way for finding exact solutions to linear partial differential equations
\begin{equation}\label{lhyp}
u_{xy} = a(x,y) u_x + b(x,y) u_y + c(x,y) u. 
\end{equation}
It is applicable if the sequence of the so-called Laplace invariants of \eqref{lhyp} is terminated by zero. This method can also be applied to the linearizations of nonlinear scalar partial differential equations   
\begin{equation}\label{hyp}
u_{xy} = F(x,y,u,u_x,u_y)
\end{equation}
as a test for Darboux integrability of \eqref{hyp} and as a way to construct higher symmetries of these equations (see, for example, \cite{AK,ZhibSok}). 

The purpose of the present notes is to demonstrate some problems that arise when we try to generalize the Laplace invariants and the method of cascade integration to the case of systems~\eqref{hyp}. That is, generally speaking, we assume that $u$ and $F$ in~\eqref{hyp} are $n$-dimensional vectors. We consider the case $n=1$ first but write almost all formulae so that they remain valid for the case $n>1$.

The mixed derivatives of $u$ can be eliminated by using~\eqref{hyp}. Therefore, we can assume without loss of generality that all local objects associated with~\eqref{hyp} are functions of the variables $x$, $y$, $u_0:=u$, $u_i := \partial^i u / \partial x^i$, $\bar{u}_j := \partial^j u / \partial y^j$. We henceforth understand functions as differential functions, i.e. they may depend on a finite number of the above variables. Let $D_x$ and $D_y$ denote the total derivatives with respect to $x$ and $y$ in virtue of~\eqref{hyp}. For any function $g$ they are defined by formulae
\begin{equation}\label{dx}
D_{x}(g) = {\frac{\partial g}{\partial x}}+ {\frac{\partial
g}{\partial u}} u_1 + \sum^{\infty
}_{i=1}\left({\frac{\partial g}{\partial u_i}} u_{i+1}
+{\frac{\partial g}{\partial
\bar{u}_i}}D^{i-1}_{y}(F)\right),
\end{equation}
\begin{equation}\label{dy}
D_{y}(g) =
\frac{\partial g}{\partial y}+\frac{\partial g}{\partial
u} \bar{u}_1 +\sum^{\infty }_{i=1}\left(\frac{\partial
g}{\partial \bar{u}_i} \bar{u}_{i+1} +\frac{\partial g}{\partial
u_i} D^{i-1}_x (F) \right).
\end{equation}

Let us consider the differential operator
\begin{equation}\label{lop}
L= D_x D_y - F_{u_x} D_x - F_{u_y} D_y - F_u.
\end{equation}
It can be written in the form
\begin{equation}\label{delim}
L= (D_y - F_{u_x}) \circ (D_x - F_{u_y}) - H_0, \qquad H_0 = F_u + F_{u_x} F_{u_y} - D_y(F_{u_y}),
\end{equation}
where the symbol $\circ$ denotes the composition of operators. Using $H_0$, $b_0=F_{u_y}$ and $L_0=L$ as starting terms, we can construct the sequences of the functions $H_i$, $b_i$ and the operators
\begin{equation}\label{lid}
L_i := (D_x - b_i) \circ (D_y - F_{u_x})  - H_{i-1} = (D_y - F_{u_x}) \circ (D_x - b_i) - H_i
\end{equation}
such that the equalities 
\begin{equation}\label{cint}
(D_x - b_i) \circ L_{i-1} = L_i \circ (D_x - b_{i-1})
\end{equation}
hold. It is easy to see that the formula
\begin{equation}\label{hi}
H_i = H_{i-1} + \left[ D_y - F_{u_x}, D_x - b_i \right]
\end{equation}
follows from~\eqref{lid}, and \eqref{cint} is equivalent to the relation 
\begin{equation}\label{bif}
(D_x - b_i) \circ H_{i-1} =  H_{i-1} (D_x - b_{i-1})
\end{equation}
for the recurrent calculation of $b_i$. Here $\left[A , B \right]$ denotes the commutator of operators $A$ and $B$. The functions $H_i$ are called \emph{Laplace $x$-invariants} of~\eqref{hyp}.

In the scalar case, equations \eqref{hi} and \eqref{bif} take the form $H_i=H_{i-1} + D_x(F_{u_x}) - D_y (b_i)$ and $b_i= b_{i-1} + D_x(H_{i-1})/H_{i-1}$. The operator equalities 
\[ L_{i-1} \circ \frac{1}{H_{i-1}} (D_y - F_{u_x}) = (D_y - F_{u_x}) \circ \frac{1}{H_{i-1}} L_i \]
allow us to construct symmetries of~\eqref{hyp} (i.e. elements of $\ker L$) from the kernel of the operator $L_r = (D_y - F_{u_x}) \circ (D_x - b_r)$ in the case $H_r=0$.

Now let us consider the case of systems. It is useful to recall some notation first. If $g$ is a scalar function and $z$ is a vector $(z^{1}$, $z^{2}$, $\dots$, ${z^{n})}^{\top}$, then by $g_{z}={{\partial g}/{\partial z}}$ we denote the row $\left( {{\partial
g}/{\partial z^{1}}}\right.$, ${{\partial g}/{\partial z^{2}}}$, $\dots$, $\left.{{\partial g}/ {\partial z^{n}}} \right)$. For any vector-valued function $G=(G^{1}$, $G^{2}$, $\dots$, ${G^{\ell})}^{\top}$, $G_z= {{\partial G}/{\partial z}}$ designates the $\ell \times n$ matrix with the rows $G^1_z$, $\dots$, $G^\ell_z$.  Taking this standard notation into account, formulae~\eqref{dx}--\eqref{bif} remain valid in the case when $u$ and $F$ in~\eqref{hyp} are $n$-dimensional vectors. Equations~\eqref{hi}, \eqref{bif} take the form
\[ H_i = H_{i-1} + D_x(F_{u_x}) + \left[F_{u_x}, b_i \right] - D_y(b_i), \] 
\begin{equation}\label{prob}
D_x (H_{i-1}) - b_i H_{i-1} +  H_{i-1}  b_{i-1} = 0
\end{equation}
in this case. But the Laplace invariants $H_i$ (as well as $b_i$) are $n \times n$ matrices now. Therefore, \eqref{prob} may have no solution $b_i$ if $\det(H_{i-1})=0$, and this solution is not unique even if it exists. Thus, we have problems with the existence and the uniqueness of the Laplace invariants in the case of systems. Some ways to deal with these problems were discussed in \cite{ZhibSok,ZhibSt}.

\medskip
{\bf Definition.} A function $w(x,y,u,u_1, \dots, u_k)$, $w_{u_k} \ne 0$, is called an \emph{$x$-integral} of system~\eqref{hyp} if $D_y(w) = 0$. The number $k$ is called the \emph{order} of this integral.
\medskip

In the scalar case, the Laplace invariants are interesting mainly in situations where \eqref{hyp} admits integrals. But the existence of an integral in the case of the systems guarantees $\det(H_r) = 0$ for some $r \ge 0$, and this leads to the aforementioned problems with the existence and the uniqueness of the Laplace invariants.

\medskip
{\bf Proposition.} {\it If system~\eqref{hyp} admits an $x$-integral of order $k$, then $\det(H_r) = 0$ for some~$r<k$.}

\medskip
This proposition was proved in \cite{StC} and then mentioned in some works. For example, it was used in \cite{ZhibSok} as one of arguments demonstrating that the direct generalization of the Laplace invariants is not completely satisfactory. But \cite{StC} is practically unavailable because it was published in Russian in a small number of copies and has no online version. The main motive for writing of the present notes is to make the proof of the above proposition freely available in English through arXiv. This proof is a modification of the reasoning that was used in \cite{AK} for the scalar case.  
\begin{proof}
For any function $g$ we can consider the differential operator
\[ g_{*}=\frac{\partial g}{\partial u} + \sum^{\infty}_{i=1} 
\left( \frac{\partial g}{\partial \bar{u}_i} D^i_y +
\frac{\partial g}{\partial u_i} D^{i}_{x} \right), \] 
i.e. $g_*$ denotes the linearization (Frech\'et derivative) of $g$.
It is not difficult to prove (see, for example, \cite{StSok}) that
\[ D_y \circ g_* - \left( D_y(g) \right)_{*} = \sum_{i=0}^p \gamma_i D_x^i \circ L, \qquad D_x \circ g_* - \left( D_x(g) \right)_{*} = \sum_{i=0}^{q} \bar{\gamma}_i D_y^i \circ L, \]
where $\gamma_i$ and $\bar{\gamma}_i$ are $n$-dimensional rows components of which are functions. Therefore,
\begin{equation}\label{fintc}
D_y \circ w_* =  \sum_{i=0}^p \gamma_i D_x^i \circ L 
\end{equation}
if $w$ is an $x$-integral.

Assume the contrary: let $\det(H_i) \ne 0$ for all $i <k$. Then for all $i \le k $ we can define the operators by the following recurrent formulae
\[ \hat{B}_0 = E, \quad \hat{B}_i=  (D_x-b_i) \circ \hat{B}_{i-1},\, i>0; \qquad
B_{-1} = E, \quad B_i=  (D_x-b_i) \circ B_{i-1},\, i \ge 0. \]
Here $E$ denotes the identity mapping ($n \times n$ identity matrix). Repeatedly using~\eqref{cint}, we obtain the relation $\hat{B}_i \circ L = L_i \circ B_{i-1}$. Since $L_i \circ B_{i-1} = (D_y - F_{u_x}) \circ B_i - H_i B_{i-1}$ follows from~\eqref{lid}, we obtain
\begin{equation}\label{dbi}
D_y \circ B_i = F_{u_x} B_i + H_i B_{i-1} + \hat{B}_i \circ L.
\end{equation}
Let $w$ be an $x$-integral of order $k$. We can rewrite its linearization in the form $w_* =  \sum_{i=0}^{k} \nu_i B_{i-1}$ and substitute this expression into~\eqref{fintc}. Then we apply the relation~\eqref{dbi} and collect the coefficients at $D_y$ and $B_i$. Since the operators $D_y$, $B_i$ and operators of the form $D_x^j \circ L$ are linearly independent, the above operations give rise to the following chain of the relations
\[ \nu_0 = 0, \qquad  \nu_1 H_0 = 0, \qquad \nu_{i+1} H_i = - \left( D_y (\nu_{i}) + \nu_{i} F_{u_x} \right),\quad 0<i<k. \]
This chain implies $\nu_i = 0$ for all $i \le k$ if $\det(H_i) \ne 0$ for all $i < k$. But $w_*=0$ contradicts the condition $w_{u_k} \ne 0$ which is contained in the definition of integrals. 
\end{proof}

\section*{Acknowledgments}
The author thanks V.V.~Sokolov for the suggestion to write this text.

\end{document}